\documentclass[12pt]{article}
\pdfoutput=1
\usepackage{graphicx,amssymb,amsmath}
\usepackage{bbm}
\usepackage{dsfont}
\newcommand{\eq}[1]{\begin{equation}#1\end{equation}}

\newcommand{\refer}[1]{(\ref{#1})}
\def\beq{\begin{equation}}
\def\eeq{\end{equation}}
\def\bea{\begin{eqnarray}}
\def\eea{\end{eqnarray}}
\def\ar{\begin{array}}
\def\ear{\end{array}}
\def\pa{\partial}

\def\nn{\nonumber}

\def\Tr{{\rm Tr}\,}

\def\ga{\gamma}
\def\Ga{\Gamma}
\def\si{\sigma}

\def\de{\delta}

\def\al{\alpha}
\def\b{\beta}

\def\la{\lambda}

\def\eps{\epsilon}

\def\ri{{\rm i}}
\def\GeV{{\rm GeV}}
\def\eV{{\rm eV}}

\def\ra{{\rightarrow}}

\DeclareFontFamily{OT1}{pzc}{}
\DeclareFontShape{OT1}{pzc}{m}{it}{<-> s * [1.10] pzcmi7t}{}
\DeclareMathAlphabet{\mathpzc}{OT1}{pzc}{m}{it}

\def\Ca{\mathpzc{a}}
\def\Cm{\mathpzc{m}}
\def\CM{\mathpzc{M}}

\def\cU{{\cal U}}

\def\a{\alpha}
\def\da{{\dot\alpha}}
\def\b{\beta}
\def\db{{\dot\beta}}

\def\ve{\varepsilon}

\def\ri{{\rm i}}

\def\cA{{\cal{A}}}
\def\cL{{\cal{L}}}

\def\cO{{\cal O}}

\def\cD{{\cal D}}

\def\cV{{\cal V}}

\def\mL{{\mathfrak{L}}}

\def\sp{\not{\!p}}

\begin{document}
\begin{center}
{\bf \Large Conformal Standard Model with\\[3mm]  an extended scalar sector}\\[1.5cm]

{\bf Adam Latosi{\'n}ski$^1$, Adrian~Lewandowski$^2$, Krzysztof A. Meissner$^2$\\[2mm] and Hermann Nicolai$^1$}

\vspace{1cm}
{{\it $^1$ Max-Planck-Institut f\"ur Gravitationsphysik
(Albert-Einstein-Institut)\\
M\"uhlenberg 1, D-14476 Potsdam, Germany\\
$^2$ Faculty of Physics,
University of Warsaw\\
Pasteura 5, 02-093 Warsaw, Poland\\

}}
\end{center}

\vspace{2cm}

{\footnotesize

{We present an extended version of the Conformal Standard Model (characterized
by the absence of any new intermediate scales between the electroweak scale
and the Planck scale) with an enlarged scalar sector coupling
to right-chiral neutrinos. The scalar potential and the
Yukawa couplings involving only right-chiral neutrinos are invariant under a
new global symmetry SU(3)$_N$ that complements the standard U(1)$_{B-L}$
symmetry, and is broken explicitly only by the Yukawa interaction, of order $\cO(10^{-6})$, 
coupling right-chiral neutrinos and the electroweak lepton doublets. We point
out four main advantages of this enlargement, namely: (1) the economy of
the (non-supersymmetric) Standard Model, and thus its observational success,
is preserved; (2) thanks to the enlarged scalar sector the RG improved one-loop effective potential
is everywhere positive with a stable global minimum, thereby avoiding the notorious
instability of the Standard Model vacuum; (3) the pseudo-Goldstone bosons resulting from spontaneous
breaking of the SU(3)$_N$ symmetry are natural  Dark Matter candidates with
calculable small masses and couplings;
and (4) the Majorana Yukawa coupling matrix acquires a form naturally
adapted to leptogenesis. The model is made perturbatively consistent
up to the Planck scale by imposing the vanishing of quadratic divergences
at the Planck scale (`softly broken conformal symmetry'). Observable consequences 
of the model occur mainly via the mixing of the new scalars and the standard model Higgs boson.
}
}

\newpage

\section{Introduction}

Experimental searches at LHC have so far not revealed any evidence of `new
physics' beyond the Standard Model (SM), and in particular no signs of low
energy supersymmetry, technicolor or large extra dimensions \cite{PDG}.
Of course, this state of affairs may change in the near future with new data,
but the possibility that there is in fact not much new structure beyond the SM
is by no means excluded. There thus remains the distinct possibility that --
apart from `small' modifications of the type suggested by the present work -- the SM may survive essentially {\em as is} all the way to the Planck scale. This prospect is further strengthened by the excellent quantitative agreement between the SM predictions and several
precision experiments that has emerged over the past decades, and which so far has not shown any deviation from SM predictions. In our view all this indicates that any `beyond the standard model' (BSM) scenario must stay as close as possible to the SM as presently understood.

The present work takes up this point of view, in an attempt to formulate a more
comprehensive and coherent scheme beyond the SM, within the general
framework proposed in \cite{MN}.  More specifically,
this is to be done in such a way that, on the one hand, the economy of the SM is maintained as much as possible, by extending it only in a very minimal way, but on the other hand, such that -- besides explaining the observed structure -- the extension solves all outstanding problems that belong to particle physics proper. The latter comprise in particular the explanation of the neutrino sector  (with light and heavy neutrinos),
the explanation of the origin of Dark Matter with suitable dark matter candidates,
and finally leptogenesis. Whereas the solution of these problems
is usually assumed to involve large intermediate scales and new heavy degrees
of freedom (GUT-scale Majorana masses, new heavy quarks to generate axion gluon
couplings, and the like) that will be difficult, if not impossible, to observe, the important point here is that we try to make do without such large scales between the electroweak and the Planck scale. This postulate entails strong restrictions that we will analyze in this work and that may be falsified by observation.
By contrast, we do {\em not} consider to belong to the realm of particle physics the problems of the cosmological constant, the origin of Dark Energy and the ultimate explanation of inflation.
Beyond their effective description in terms of scalar fields, these are here assumed to involve quantum gravity in an essential way, whence their solution must await the advent of a proper theory of quantum gravity.

The crucial assumption underlying the present work, and the defining property 
of the term `Conformal Standard Model' (CSM)\footnote{In order to avoid an unnecessary
  proliferation of names, we have decided to use this name for the whole class of models
 satisfying the stated requirements.}, is conformal symmetry, albeit in a `softly broken'
form, and consequently  the absence of any new scales intermediate between 
the electroweak scale and the Planck scale. This basic assumption is motivated on the one hand by the absence of any direct evidence of such intermediate mass scales, and on the other hand by the `near conformality' of the SM, that is, the fact that the SM {\em is} classically conformally invariant, {\em except} for the the explicit mass term in the scalar potential introduced to trigger spontaneous symmetry breaking. In previous work we have formulated a scenario which attempts to exploit this fact, and thus to explain the stability of the electroweak scale as well as the supposed absence of large intermediate scales, by imposing classical conformal symmetry as a basic symmetry. Importantly, we thus do not rely on low energy supersymmetry
to explain the stability of the electroweak scale. In \cite{MN} the Coleman-Weinberg 
mechanism \cite{CW} was invoked to provide a quantum mechanical source of conformal symmetry breaking, but more recently we have adopted a variant of this scheme, by allowing for explicit mass terms, but with the extra restriction of vanishing quadratic divergences in terms of bare parameters at the Planck scale, in a realization of what we call `softly broken conformal symmetry' \cite{SBCS}. With either realization there is then only one scale other than the Planck scale in the game; this scale, which should be tiny in comparison with the Planck scale, is here assumed to be $\cO$(1)\,TeV. The challenge, then, is to accommodate within such a scenario  all observed SM phenomena and, in particular, the considerable differences in scales observed in the SM.
To these requirements we add the triple conditions of {\em perturbative consistency} (absence of Landau poles up to the Planck scale $M_{Pl}$), of {\em lower boundedness of the RG improved one-loop effective potential $\ \mathcal{V}_{\rm eff}^{RGI}(\varphi)$}, and finally, of {\em vacuum stability} (the electroweak vacuum should remain the global minimum of 
$\ \mathcal{V}_{\rm eff}^{RGI}(\varphi)$ in the region $|\!| \varphi |\!| \lesssim M_{Pl}$). 
It is a non-trivial check on our assumptions that there do
exist parameter values satisfying all of these constraints.

Accordingly, we present in this paper a slightly modified version of the model proposed in~\cite{MN,SBCS}, with the aim of working towards a more comprehensive scenario of BSM physics.
The modification consists in an enlargement of the scalar sector that couples to the right-chiral neutrinos, and the introduction of a new global SU(3)$_N$ symmetry
acting only on the right-chiral neutrinos and the new scalar fields. This symmetry
is assumed to be spontaneously broken, giving rise to several Goldstone bosons.
The latter are converted to pseudo-Goldstone bosons by the one-loop corrections
induced by the Yukawa interaction coupling right-chiral neutrinos and the electroweak  lepton doublets, which is the only term in the Lagrangian that breaks SU(3)$_N$ explicitly.
Besides preserving the economy of the (non-supersymmetric) SM, this version of the
CSM  comes in particular with the following advantages: (1) the pseudo-Goldstone
bosons resulting from spontaneous symmetry breaking can in principle serve as
Dark Matter candidates with calculable small masses and couplings, and (2) the Majorana Yukawa coupling matrix dynamically acquires a form naturally adapted to leptogenesis via the mechanism proposed and investigated in \cite{PU}. Furthermore, there remains the possibility that a certain linear combination of the pseudo-Goldstone bosons may be identified with the axion
required for the solution of the strong CP problem\,\footnote{In our previous work \cite{LMN}
it was wrongly claimed that the Majoron can become a
pseudo-Goldstone boson. The error in that argument, which was based
on a rather involved three-loop calculation, was uncovered
thanks to the new technology developed in \cite{Lat}, which shows that only
fields orthogonal  to the identity in the matrix of Goldstone fields can become
pseudo-Goldstone bosons, cf. (\ref{Amass}) and section~2.4 for details.}.

Finally, we briefly discuss two natural extensions of our main model, namely 
first, the possibility of gauging U(1)$_{B-L}$, and secondly  a further enlargement 
of the scalar sector that changes the breaking pattern of the SU(3)$_N$ symmetry.

\subsection*{Related and previous work}
We should note that there is a substantial body of work along similar lines as proposed
here, and we therefore briefly recall and comment on some related proposals.
The idea of exploiting the possible or postulated absence of intermediate scales in order
to arrive at predictions for the Higgs and top quark masses was already considered 
in \cite{FN}. However, it appears that the actual values of the SM parameters 
with only the standard scalar doublet cannot be reconciled with the stability of the electroweak vacuum over the whole range of energies up to the Planck mass  (see \cite{Nielsen} and \cite{Turok} for a more recent re-assessment of this scenario). The possible importance of conformal symmetry 
in explaining the electroweak hierarchy was already emphasized in \cite{Bardeen}. 
More recently, there have been a number of approaches proceeding from the assumption of conformal symmetry, in part based on the Coleman-Weinberg mechanism, as in \cite{CW4,CWzz}
and \cite{CW1,CW2}. The latter papers discuss in particular 
aspects of neutrino physics in conformal 
theories; see also  \cite{CW3,CWzzz,CWz4} for a discussion of the phenomenology of such models. The idea that radiative electroweak symmetry breaking is triggered by a new U(1)$_X$ gauge boson without direct couplings to SM particles was introduced in \cite{Hemp} and reconsidered more recently in \cite{NoHemp1, NoHemp2}. An extension to the case of
an SU(2)$_X$ gauge group was proposed in \cite{CaRa}, raising the interesting possibility 
of a (non-abelian) spin-1 Dark Matter particle \cite{Hambye}.  
Conformal models with local $(B\!-\!L)$ symmetry have been investigated 
in \cite{CW5a,CW5}, exploiting the same mechanism as in \cite{CW} to stabilise
the radiatively generated vacuum. For these gauge groups the phenomenology 
was recently reanalyzed in \cite{phenoCW}. The unavoidable mixing between multiple 
U(1) factors \cite{Chank} was included in the study of the U(1)$_{B\!-\!L}$ case \cite{CWa},
which also addresses the issue of vacuum stability with U(1)$_{B-L}$ gauging.
RG improved effective potentials and their applications in the conformal context were 
considered in \cite{CW6,CW7,CWb}. The possibility that all scales are generated 
dynamically was also considered from another point of view in \cite{noCWxx, noCWyy}. 
Furthermore it has been pointed out in \cite{SW} that the vanishing of the SM scalar self-coupling and the associated $\beta$-function at the Planck scale could be interpreted as evidence for a hidden conformal symmetry at that scale (and also for asymptotic safety); this proposal 
is in some sense the opposite of the present scenario, where conformal symmetry is 
assumed to be relevant {\em below} the Planck scale. Among the non-supersymmetric 
attempts at a comprehensive approach to BSM physics the so-called $\nu$MSM model 
of \cite{Shap} has been widely discussed; this model is somewhat related to the  
present work in that it is also based on a minimal extension of the SM, but differs in 
other aspects (for instance, in trying to incorporate inflation, with the Higgs boson 
as the inflaton).  Other non-supersymmetric proposals with `new physics' in the 
range accessible to LHC include the twin Higgs models \cite{a1}, minimal models 
with fermionic \cite{a2,a3,a4} or scalar Dark Matter \cite{a5}, as well as other interesting 
possibilities, e.g. \cite{a6,a7,a8,a9,a10,a11,a12}. 
Here we will have nothing to say about supersymmetric models,
which are characterised by more than just minimal additions to the SM, and where there 
is a vast literature, see e.g. \cite{Buchmueller} for a recent overview and bibliography.

\section{Basic features of the model}
To present our point of view in as clear a manner as possible this paper is
structured in line with our basic assumptions, which concern in particular
\begin{itemize}
\item Scalar sector
\item Fermionic sector
\item Pseudo-Goldstone bosons and their couplings
\end{itemize}
and which we will discuss in this section. In the following section we will discuss
the constraints that self-consistency and compatibility with the SM and
other data impose on the model and its parameters. Possible checks (that 
could in principle falsify our approach) are also discussed there, as well as
possible signatures that may discriminate the present proposal from
other proposals.

\subsection{Scalar Sector}\label{Sec:ScalarSec}

Although full confirmation is still pending, there is good evidence
that the SM Higgs boson does not distinguish between different families
(generations) \cite{PDG}. Consequently, its different couplings to the SM fermions
are entirely due to the different Yukawa coupling matrices, implying for instance
that the Higgs couplings to quarks and leptons are directly proportional to their masses.
It would therefore seem natural to assume that possible extensions of the scalar
sector to include Majorana-like couplings to the right-chiral neutrinos should also proceed through a `family blind' electroweak singlet scalar $\phi$ whose vacuum expectation value generates the usual Majorana mass term required for the seesaw mechanism, with an appropriate Majorana-type Yukawa coupling matrix $Y^M_{ij}$, and this path has been followed mostly in past work. By contrast, we here wish to explore an alternative scenario relaxing this assumption, and to point out several advantages that come with making the extended scalar sector sensitive to the family structure of right-chiral neutrinos. These concern in particular the appearance of pseudo-Goldstone bosons that are natural Dark Matter candidates, with calculable small masses and couplings. Furthermore, thanks to the new scalar fields, the much advertised instability  of the Higgs coupling and the one-loop effective potential in the (un-extended) SM (see e.g. \cite{Degrassi,Buttazzo,HKO,Branchina} for a recent  discussion) can be avoided without great effort.

Accordingly, the main new feature of our model in comparison with the SM is its enlarged scalar sector, while there is no corresponding enlargement in the fermionic sector, other than the {\it ab initio} incorporation of right-chiral neutrinos (see below). The scalar sector is assumed, on the one hand, to allow for a Majorana mass matrix for the right-chiral neutrinos to be  generated by spontaneous symmetry breaking, and with a breaking pattern adapted to leptogenesis, and on the other to allow for the existence of very light pseudo-Goldstone bosons that can serve as natural dark matter candidates. The appearance of extra scalar degrees of freedom is a common feature of many proposed extensions
of the SM, and in particular, of supergravity and superstring scenarios.
A distinctive feature of the present scheme is that the new scalars, while
carrying family indices, are otherwise `sterile',  except for those
scalars that mix with the standard Higgs boson; as we will explain below this can lead to new experimental signatures, different from low energy supersymmetry and other scenarios where extra scalars carry electroweak or strong charges. The assumed sterility safeguards principal successes of the SM, in particular the absence of FCNC.
While it might appear desirable to also extend the family structure of the scalars
to the quark and lepton sector, our assumption of    `near conformality' seems difficult to reconcile with the existence of scalars relating different generations of quarks and leptons: by softly broken conformal invariance these would have to have relatively low masses, and thus conflicts would with SM data would be inevitable. In this respect,  the situation is different in GUT-type scenarios,  where such extra scalars can in principle be made sufficiently heavy so as to avoid any direct conflict with observation. However, even in that context, fully consistent models with family sensitive scalars seem hard to come by, and we are not aware of a single example of a model of this type that works all the way (see, however, \cite{Ramond} and references therein for a recent attempt to explain the observed hierarchy of quark masses in terms of discrete subgroups of a family symmetry SU(3)).

A new feature in comparison with \cite{MN} is thus that the scalars coupling
to the right-chiral neutrinos are assumed to admit a family-type symmetry SU(3)$_N$
that complements the standard U(1)$_{B-L}$ symmetry. This new symmetry is 
broken explicitly by the Dirac-Yukawa couplings $Y^\nu$;  importantly, the latter 
are very small (of order $\cO(10^{-6})$). Accordingly, we introduce a complex 
scalar sextet $\phi_{ij}= \phi_{ji}$ (with family indices $i,j,...$) which are `blind' to the SM gauge
symmetry, hence sterile. This sextet replaces the standard Majorana mass
term triggered by a family singlet scalar $\phi$ according to
\beq
\langle\phi\rangle \, Y^M_{ij}  \;\; \longrightarrow \;\;  y_M \, \langle\phi_{ij}\rangle,
\eeq
and similarly for the associated Majorana-type  Yukawa couplings.
With the usual Higgs doublet $H$ the scalar field Lagrangian is
\beq
\mathfrak{L}_{scalar}=-(D_\mu H)^\dagger(D^\mu H)-\Tr\!(\partial_\mu\phi^*\partial^\mu\phi)
-\cV({H}, \phi).
\eeq
The potential is
\bea
\cV({H}, \phi) &=&  m_1^2 \, {H}^\dag{H} +  m_2^2 \, \Tr\!(\phi  \phi^*)
    +    \, \la_1 \,({H}^\dag{H})^2  \,    \\[2mm]
 && \!\!\!\!\!\!\!\!\!\!\!\!\!\!\!\!\!\!\!\!\!\!\!\!    + \,
 2 \la_3 \, ({H}^\dag{H}) \Tr\!( \phi \phi^*) \,  +  \, \la_2  \left[\Tr\!(\phi\phi^*)\right]^2
\,  + \,  \la_4 \, \Tr\! (\phi \phi^* \phi \phi^*),\nn
\label{scpot}
\eea
where all coefficient are real (traces are over family indices). This potential is manifestly invariant under
\beq\label{U3N}
\phi (x) \quad \ra \quad U \phi(x) U^T \; ,
\quad\quad\quad U\in {\rm U}(3).\\
\eeq
The scalar fields $\phi_{ij}$ are inert under the usual SM symmetries, unlike the Higgs doublet $H$.

There are three different cases that ensure positive definiteness of the 
quartic part of the classical potential
\begin{itemize}
\item $\lambda_1,\lambda_2,\lambda_4>0,\ \ \lambda_3>-\sqrt{\lambda_1(\lambda_2+\lambda_4/3)};$
\item $\lambda_4<0,\ \  \lambda_1>0,\ \ \lambda_2>-\lambda_4, \ \ \lambda_3>-\sqrt{\lambda_1(\lambda_2+\lambda_4)}; $
\item $\lambda_2<0,\ \  \lambda_1>0,\ \ \lambda_4>-3\lambda_2, \ \ \lambda_3>-\sqrt{\lambda_1(\lambda_2+\lambda_4/3)}. $

\end{itemize} 

One of these conditions has to hold {\em for all scales} between the electroweak  and the Planck scales to avoid the problem of vacuum instability, and thus to overcome one of the main
open problems of the SM in its current form. More concretely, we will require them to hold for the running couplings $\lambda_i(\mu)$ over this whole range when
these are evolved with the $\beta$-functions (\ref{beta}).

Assuming the following values of the mass parameters
\beq
m_1^2=-2\lambda_{1}v_{H}^{2}-6\lambda_{3}v_{\phi}^{2},\qquad
m_2^2=-2\lambda_{3}v_{H}^{2}-\left(6\lambda_{2}+2{\lambda_{4}}\right)v_{\phi}^{2},
\eeq
(and thus parametrising them directly in terms of the positive 
parameters $v_H$ and $v_\phi$)
it is straightforward to show that the global minimum of the potential takes the form\footnote{Thus, $v_H$ and $v_\phi$ are the expectation values of the complex fields (we here drop the customary factor $1/\sqrt{2}$).}
\beq\label{vevs}
\langle {H} \rangle = \left(\begin{array}{c} 0 \\[1mm]
                                                        v_H  \end{array}\right), \qquad
\langle \phi \rangle =
\mathcal{U}_0\left(\begin{array}{ccc} v_\phi&0&0 \\[1mm]  0&v_\phi&0 \\[1mm]   0&0&v_\phi  \end{array}\right)\mathcal{U}_0^{T},
\qquad \mathcal{U}_0\in {\rm U}(3),
\eeq
provided that (in addition to the above positivity conditions) the following inequalities are also satisfied
\eq{\nn
\lambda_1 \left\{{\lambda}_{2}+\frac{{\lambda}_{4}}{3}\right\}-\lambda^2_3 >0, \qquad
\lambda_4>0.
}
At the classical level the U(3) matrix $\cU_0$ remains undetermined. %
The explicit breaking of SU(3)$_N$ symmetry to be discussed below
will, however, lift this degeneracy and produce a `vacuum alignment'  with 
$\cU_0\neq \mathbbm{1}$ according to
\cite{DASHEN},  and also introduce small corrections that will lift the
degeneracy of  eigenvalues in $\langle\phi_{ij}\rangle$.

A second motivation for the replacement of a single complex scalar by a sextet is the following. Because the SU(3)$_N$ invariance is assumed to be broken both spontaneously and explicitly  (by the Yukawa interaction coupling right-chiral neutrinos to the lepton doublets via the matrix $Y^\nu_{ij}$, see (\ref{L'}) below)
there exist various light particles, i.e. (pseudo-)Goldstone bosons.
It is a general result that the manifold of Goldstone bosons $\cal{M}$
is the quotient of the symmetry group by the symmetry of the vacuum. For (\ref{vevs}) the residual symmetry is SO(3), and therefore
\beq
\label{MGoldstone}
{\cal M} = {\rm U}(3)\big/ {\rm SO}(3) \, \equiv \,
 {\rm U}(1)_{B-L} \,\times\,  {\rm SU}(3)_N\big/{\rm SO}(3),
\eeq
whence there are altogether six (pseudo-)Goldstone bosons in our model. One of them is the \emph{genuine} Goldstone boson associated with the exact U(1)$_{B - L}$ symmetry (so
we can take out the U(1) factor).

After the symmetry breaking we have as usual the real Higgs field $H_0$  
\beq
H(x)= \left(\begin{array}{c} 0 \\[1mm]
                        v_H  +\frac{1}{\sqrt{2}}H_0(x)\end{array}\right),
\eeq
{(in the unitary gauge)} 
while a convenient parametrisation of the coset space $\cal M$ is given by
\beq\label{ReDef1}
\phi(x)=\mathcal{U}_0\,e^{\ri \tilde{A} (x) }\,(v_\phi+\tilde{R}(x))\,  e^{\ri \tilde{A}(x)}\,
\mathcal{U}_0^{\rm T},
\eeq
where $\tilde{A}_{ij}$ and $\tilde{R}_{ij}$ are real symmetric matrices. The trace part of
\beq\label{calG}
\mathcal{G}(x)\equiv\mathcal{U}_0\tilde{A}(x)\mathcal{U}_0^{\dagger},
\eeq
is the $(B - L)$ Goldstone boson $\Ca(x)$ that remains a Goldstone boson even when 
the U(3) symmetry is broken, while the traceless part of
$\mathcal{G}(x)$ yields the five Goldstone bosons that will be converted to  pseudo-Goldstone bosons. In accordance with the decomposition ${\bf 6} \ra {\bf 1} \oplus {\bf 5}$ under the residual SO(3) symmetry we can thus write
\beq\label{A}
\tilde{A}_{ij} (x) = \frac{1}{2\sqrt{6}v_\phi}\Ca(x) \delta_{ij} \, + \, A_{ij}(x)  \;\;,
\qquad \Tr A(x) = 0,
\eeq
with
\beq\label{G}
A_{ij} \equiv \frac1{v_\phi} G(x) \equiv \frac{1}{4v_\phi} {\sum_a}' G_a\la^a_{ij},
\eeq
where the restricted sum is only over the five {\em symmetric} Gell-Mann  matrices 
(with the standard normalization $\Tr(\la^a \la^b)=2\delta^{ab}$)
and  where the real fields $\Ca(x)$ and $G_a(x)$ are canonically normalized.
The matrix $\tilde{R}_{ij}(x)$ likewise can be split into a trace and a traceless part, {\it viz.}
\beq\label{R}
\tilde{R}_{ij}(x) = \frac{1}{\sqrt{6}}r(x) \delta_{ij} \, + \frac{1}{2}\, {\sum_a}' R_a\la^a_{ij} (x).
\eeq

Because the new scalars are thus only very weakly coupled to the remaining SM fields, the main observable effects are due to the mixing between the standard Higgs boson and the new scalars.
In fact, the five modes $R_a$ are already the proper mass eigenstates with eigenvalues
\beq
M_R^2=4\lambda_4 v_\phi^2,
\eeq
The field $r$ can mix with $H_0$ and the combined mass matrix for the fields $(H_0,r)$ reads
\beq
\mathcal{M}^2
=\left(\begin{array}{cc}
4\lambda_1 v_H^2 & 4\sqrt{3} \lambda_3 v_H v_\phi
\\
4\sqrt{3} \lambda_3 v_H v_\phi  & 4(3\lambda_2+\lambda_4) v_\phi^2
\end{array}\right),
\eeq
and determines the mass eigenstates $h_0$ and $h'$
\beq\label{hh'}
h_0 \,=\, \cos\beta\, H_0 \, + \, \sin\beta \, r  \;\; , \quad
h' \, = \,  - \sin\beta \, H_0 \, + \, \cos\beta \, r,
\eeq
with the mixing angle $\beta$. We identify the lighter of the two mass eigenstates $h_0$ with the observed Higgs boson, with $M_{h_0}\approx 125\GeV$.  The mixing will lead to a second resonance associated with $h'$, which
is one of the main predictions of the present model. This resonance
should be rather narrow because of the factor $\sin^2\beta$ \cite{MN325}. It will have the same decay channels to the SM particles as the standard Higgs boson (hence look like a `shadow Higgs'), but depending on the actual mass of $h'$,  there may also be other decay channels which could broaden the resonance. We will return to this point below.

The possibility of further extension of the Higgs sector in the framework of a U(3) symmetric
scalar sector is considered in the Appendix.

\subsection{Fermionic sector}

With right-chiral neutrinos, the SM comprises altogether 48 fundamental spin-$\frac12$ degrees of freedom, in three generations (families) of 16 fermions each. It is one of our basic assumptions that there are no other spin-$\frac12$ degrees of freedom.\footnote{The occurrence of 16 fermions in one generation is often
  interpreted as strong evidence for an underlying SO(10) GUT symmetry. However,
  apart from the fact that SO(10) cannot explain the origin of the family replication,
  there may be alternative explanations. In particular, 48 = 3$\,\times\,$16 is also the number of physical spin-$\frac12$ fermions in maximally extended ($N=8$)
  supergravity remaining after complete breaking of supersymmetry.
  See \cite{MNN8} for a fresh look at this coincidence.}
This assumption is mainly motivated by observation, that is, the complete
lack of evidence so far of such new fermionic degrees of freedom at LHC.
In fact, already the LEP experiment had produced strong evidence that there
exist  only three generations, so any extra spin-$\frac12$ fermions beyond the
known quarks and leptons would have to be either sterile, or otherwise appear as
heavy fermionic superpartners of the known SM bosons (thus not implying the
existence of new families of fermions).

We here concentrate on the Yukawa part of the extended CSM Lagrangian,
referring to \cite{Weinberg,EPP,Pok} for the complete SM Lagrangian and its properties.
With the above assumptions concerning the fermionic sector and the new scalar
sextet introduced in the foregoing section, we can write down right away the most general Yukawa couplings: the Higgs doublet couples in the usual way, while $\phi_{ij}$ couples only to the right-chiral neutrinos.
Accordingly, the complete Yukawa part of the Lagrangian is~\footnote{We will make
  consistent use of Weyl (two-component) spinors throughout, see e.g. \cite{LMN} for our conventions, as we have found them much more convenient than 4-spinors in
  dealing with the intricacies of the neutrino sector.}
\bea\label{LY}
\mL_{\rm Y}  &=&
\big\{- Y_{ij}^E {H^\dagger}\!{L}^{i\alpha} E^j_\alpha
- Y_{ij}^D {H}^{\dagger}\!{Q}^{i\alpha} D^j_\alpha - Y_{ij}^U {H}^T\!\ve{Q}^{i\alpha} U^j_\alpha \nn\\
&& - \, Y_{ij}^\nu {H}^T\!\ve{L}^{i\alpha} N^j_\alpha -
 \frac12{y_M} \phi_{ij} N^{i\alpha} N^j_\alpha\big\} \, +\, {\rm h.c.}
\eea

Here $Q^i_\alpha$ and $L^i_\alpha$ are the left-chiral quark and lepton doublets,
$\bar{U}^{i\dot{\alpha}}$ and $\bar{D}^{i\dot{\alpha}}$ are the right-chiral up- and down-like quarks, while $\bar{E}^{i\dot{\alpha}}$ are the right-chiral electron-like leptons, and $\bar{N}^{i\dot{\alpha}}$ the right-chiral neutrinos; the family indices $i,j=1,2,3$ as well as $SL(2,\mathbb{C})$ indices are written out explicitly. Classically, the full SM Lagrangian is invariant under lepton number
symmetry  U(1)$_L$ as well as under the usual baryon number symmetry U(1)$_B$; these two U(1) symmetries combine to the anomaly free U(1)$_{B-L}$ symmetry which is hence preserved to all orders.

The main new feature of our model is that the right-chiral neutrinos transform under the previously introduced symmetry SU(3)$_N$ according to
\beq
N^i(x) \quad \rightarrow \quad (U^*)^i{}_j N^j(x),
\eeq
whereas all other SM fermions are inert under this symmetry.\footnote{Strictly speaking,
   we should therefore use two different kinds of family indices, one for the usual quarks and leptons, the other for the right-chiral neutrinos. We will refrain from doing so in order to keep the notation simple.}
This reflects the essential difference in our model between the quarks and leptons on the one hand, where the Yukawa couplings are given by numerical matrices, and the right-chiral neutrinos on the other, where the \emph{effective} couplings are to be determined as vacuum expectation values of sterile scalar fields. The SU(3)$_N$ symmetry is thus broken explicitly only by one term in (\ref{LY}), namely the interaction
\beq\label{L'}
\mL_{\rm Y}' =  -Y_{ij}^\nu {H}^T\!\ve{L}^{i\alpha} N^j_\alpha \, + \, {\rm h.c.},
\eeq
coupling the lepton doublet and the right-chiral neutrinos. Consequently, (\ref{L'}) is the only term in the SM Lagrangian by which right-chiral neutrinos communicate with the rest of the SM, and hence will play a key role in the remainder. We repeat that this interaction {\em does} preserve U(1)$_{B-L}$.
The numerical matrix $Y_{ij}^\nu$ here must be assumed very small [with entries
of order $\cO(10^{-6})$],  in order to explain the smallness of light neutrino masses via the see-saw mechanism \cite{Min,seesaw,Yan,Mo2}  with TeV scale heavy neutrinos.

The neutrino masses emerge upon spontaneous symmetry breaking in the usual
way, and thus depend on the matrices $m_{ij}$ and $M_{ij}$ defined by the
vacuum expectation values of the corresponding scalar fields, {\it viz.}
\beq\label{MY}
M_{ij}=y_M\langle\phi_{ij}\rangle, 
\eeq
and
\beq\label{mY}
m_{ij} = \, Y^\nu_{ij} \, v_H.
\eeq
Given these matrices, the (squared) masses of the light neutrinos are then determined as the eigenvalues of the following matrices (see e.g. \cite{LMN} for a derivation), namely
\beq
{\bf m}_\nu^2=\Cm^\dagger \Cm,\ \ \ \ \ \ \ \mbox{with $\;\Cm \equiv m M^{-1}m^T+\ldots$},
\label{MassF}
\eeq
for the light neutrinos, and
\beq\label{MN}
{\bf M}_N^2=\CM^\dagger \CM,\ \ \ \ \ \ \
\mbox{with $\;\CM \equiv M+\frac12 m^T m^*M^{-1}+\frac12 M^{-1} m^\dagger m +\ldots$},
\eeq
for the heavy neutrinos. These formulas generalize the well-known seesaw
mass formulas \cite{Min,seesaw,Yan,Mo2}.
Assuming $m\sim 100$ keV and $M\sim 1$ TeV we get light neutrinos with masses of order $0.01$ eV, and heavy neutrinos with masses of order 1 TeV. The mass eigenvalues are furthermore constrained by the known mass differences $\de m_\nu^2$.

We conclude this section by giving the neutrino propagators derived in \cite{LMN}
for the case when $M_{ij}$ is given by (\ref{MY}) with \refer{vevs}. %
With a proper change of basis in the space of right-chiral neutrinos we can assume, that $M_{ij}=M \delta_{ij}$ with a positive parameter $M$ (this change will also modify $m$, see below).   
Moreover, because the effects we are looking for depend on the small matrix $m_{ij}$
we can simplify the expressions further by expanding in powers of $m_{ij}$. Up
to and including terms $\cO(m^3)$ this gives  (suppressing family indices)
\bea
\langle \nu_\a \bar\nu_\db \rangle &=& -\ri \frac{\sp_{\a\db}}{p^2}\big(1-m^*\cD(p)^*(p^2+m^\dag m)m^T\big) = \nn\\[2mm]
&=& -\ri \sp_{\a\db}\Big(\frac{1}{p^2} - \frac{m^*m^T}{p^2(p^2+M^2)} + \dots\Big),
\nn\\[2mm]
%
\langle N_\a \bar N_\db \rangle &=& -\ri \sp_{\a\db} \cD(p) (p^2+m^Tm^*) = \nn\\[2mm]
&=& -\ri \sp_{\a\db} \Big(\frac{1}{p^2+M^2} \, - \,
\frac{m^\dag m}{(p^2+M^2)^2}+\frac{M^2m^Tm^*}{p^2(p^2+M^2)^2}
\, + \dots\Big),
\nn\\[2mm]
\langle \nu_\a \nu_\b \rangle &=& \ri \eps_{\a\b} M m^*\cD(p)^*m^\dag = \nn\\[2mm]
&=& \ri \eps_{\a\b} \Big(\frac{Mm^*m^\dag}{p^2(p^2+M^2)} + \dots\Big), \nn\\[2mm]
\langle N_\a N_\b \rangle &=& -\ri \eps_{\a\b} M p^2 \cD(p) = \nn\\[2mm]
&=& -\ri \eps_{\a\b} \Big(\frac{M}{p^2+M^2}
\, -\, \frac{M (m^\dagger m + m^Tm^*)}{(p^2+M^2)^2}
\, +\dots\Big),
\nn\\[2mm]
\langle \nu_\a \bar N_\db \rangle &=& \ri \sp_{\a\db} M m^*\cD(p)^* = \nn\\[2mm]
&=& \ri \sp_{\a\db} \Big(\frac{M m^*}{p^2(p^2+M^2)}
\, -\, \frac{M m^* (m^Tm^* + m^\dagger m)}{p^2(p^2+M^2)^2}
\, +\dots\Big), \nn\\[2mm]
\langle \nu_\a N_\b \rangle &=& -\ri \eps_{\a\b} m^*\cD(p)^*(p^2+m^\dag m) = \nn\\[2mm]
&=& -\ri \eps_{\a\b} \Big(\frac{m^*}{p^2+M^2} \, - \,
\frac{m^*m^Tm^*}{(p^2+M^2)^2}+\frac{M^2m^*m^\dagger m}{p^2(p^2+M^2)^2}
\, + \dots\Big),
\label{expprops}
\eea
where
\bea
&\sp_{\a\db} = p_\mu \si^\mu_{\a\db}, \qquad \si^\mu = (\mathbbm{1},\si^i), \nn \\
&\cD(p) \,\equiv \, \big[(p^2+m^Tm^*)(p^2+m^\dag m)+p^2 M^2\big]^{-1} = \cD(p)^T.
\eea
In evaluating the Feynman integrals we should keep in mind that expressions
containing the matrix $m_{ij}$ can originate both from this expansion as well
as from the interaction vertex  (\ref{LYnew}) below.

\subsection{Pseudo-Goldstone bosons}

As we already pointed out, besides the Majoron, there appear five Goldstone bosons.
The latter are converted to pseudo-Goldstone bosons via radiative corrections that originate from the Yukawa term (\ref{L'}).
To make all this more explicit we need to parametrize the Goldstone manifold $\cal M$ in (\ref{MGoldstone}). %
To this aim, we first separate off the (pseudo-)Goldstone modes by means of the
formula (\ref{ReDef1}). According to (\ref{A}) we can then split $\tilde A_{ij}(x)$ into a trace part and the rest. As we will see below, because of the
explicit breaking of SU(3)$_N$, and hence also its SO(3) subgroup, induced by the
Yukawa couplings $Y^\nu_{ij}$, the five Goldstone fields contained in $A_{ij}(x)$
will actually acquire very small masses, and  thus metamorphose
into pseudo-Goldstone bosons.

To proceed it is convenient to eliminate the pseudo-Goldstone boson fields from
the Majorana Yukawa coupling $\propto \phi NN$ by absorbing them into the
right-chiral neutrino spinors 
\beq
N^i_\alpha(x) = (\mathcal{U}_0^*e^{-\ri \tilde A(x)}\mathcal{U}_0^T)^i{}_j  \tilde{N}^j_\alpha(x),
\eeq
where we have included the (constant) `vacuum realignment matrix'  $\cU_0$ that is implicitly determined by requiring absence of tadpoles (or equivalently,
$\langle A\rangle = 0$ for the vacuum of the one-loop corrected effective potential, see below).
For the remaining SM fermions there is a similar redefinition only involving the
Majoron $\Ca(x)$. After this redefinition the Goldstone modes  only appear in
the Dirac-Yukawa coupling (\ref{L'}) and via derivative couplings of the type
$\partial_\mu A \bar{f}\gamma^\mu \frac{1+\ga_5}{2}f$. The only {\em non-derivative}
couplings of the pseudo-Goldstone fields are thus given by
\bea\label{LYnew}
\mL_{\rm Y}' &=&  - (Y^\nu\, \mathcal{U}_0^*e^{-\ri A(x)}\mathcal{U}_0^T)_{ij} {H}^T\! \ve \tilde{L}^{i\alpha} \tilde N^j_\alpha\, + \, {\rm h.c.} \nn\\[2mm]
 &=& -\, v_H (Y^\nu\, \mathcal{U}_0^*e^{-\ri A(x)}\mathcal{U}_0^T)_{ij}
 \tilde{\nu}^{i\alpha}\tilde N^j_\alpha \, + \, {\rm h.c.} \, + \, \cdots .
\eea
The Majoron $\Ca(x)$ has disappeared from the above interaction term because of the
accompanying redefinitions of the left-chiral leptons, in accordance with exact
U(1)$_{B-L}$ symmetry (thus, $\Ca(x)$ has  {\em only} derivative couplings).
Even though the interaction (\ref{LYnew}) now looks non-renormalizable, it is, of course, not. However, in order to recover renormalizability in this `picture' one
must expand the  exponential as appropriate. For instance, at one loop we will have to take into account both linear and quadratic terms in $\tilde A(x)$ when computing mass corrections, see below.

At this point we can also absorb the vacuum realignment matrix $\cU_0$ into
a redefinition of the Yukawa couplings. For this purpose we redefine the right-chiral neutrino fields once again
\beq
\tilde{N}^j_\alpha(x) \equiv \left(\frac{y_M^*}{|y_M|}\right)^{1/2}(\mathcal{U}_0^*)^j_{\ i}\widehat{N}^i_\alpha(x)\, ,
\eeq
so that, in terms of new fermion fields, the vertex \refer{LYnew} reads
\bea\label{LYnewSimplified}
\mL_{\rm Y}' &=& -
v_H (\hat{Y}^\nu\, e^{-\ri A(x)})_{ij} \tilde{\nu}^{i\alpha}\hat N^j_\alpha \, + \, {\rm h.c.} \, + \, \cdots,
\eea
with the redefined Yukawa coupling matrix
\beq
\hat{Y}^\nu=\left(\frac{y_M^*}{|y_M|}\right)^{1/2} Y^\nu\mathcal{U}_0^*.     \label{newYnu}
\eeq
The presence of a non-trivial vacuum realignment matrix $\cU_0$ entails the
following redefinition of the Dirac and Majorana mass matrices
\beq
\hat{m}=\hat{Y}^\nu\, v_H,\qquad
\hat{M}=|y_M| v_\phi \mathbbm{1}.
\eeq
so the redefined Majorana mass matrix is diagonal.

For the calculation of the radiative correction we employ the neutrino matrix propagators listed in (\ref{expprops}), with $M\equiv|y_M| v_\phi$ and the replacement $m\mapsto\hat{m}$. While the original potential did not depend on $A(x)$ at all, this vacuum degeneracy is lifted at one loop due to the interactions induced by the term (\ref{LYnew}) in the effective potential.
The result can be expanded in powers of $A$, but we are interested here only in the linear and quadratic terms. %
There is a finite contribution to the term linear in $A_{ij}$, which is proportional to
\beq
\int \frac{d^4p}{(2\pi)^4} \frac{M^2}{p^2(p^2+M^2)^2}\,
\Tr \Big( \big[ \hat{m}^\dag  \hat{m}\,,\, \hat{m}^T \hat{m}^*\big] \, A \Big),
\label{Alin}
\eeq
and comes from the tadpole diagram, using the $\langle\nu_\a N_\b\rangle$
and $\langle\bar\nu_\da \bar N_\db\rangle$ propagators from (\ref{expprops}), with
one factor $m$ from the vertex and the other factors containing $m$ from the propagators.
{\em Importantly, there are neither quadratic nor logarithmic divergences.}
To identify the true vacuum we now require absence of tadpoles \cite{DASHEN}, or equivalently, the vanishing of the linear term above. This amounts to choosing the 
vacuum realignment matrix $\cU_0$ in such a way that  the commutator in (\ref{Alin}) 
vanishes. Equivalently, we demand the matrix
\beq
\hat{m}^\dag  \hat{m} =\cU_0^T (m^\dag m) \cU_0^*,
\eeq
to be real; the requisite re-alignment matrix $\cU_0$ a matrix always exists because 
$m^\dag m$ is hermitean. Consequently,
\beq\label{mm}
\big[ \hat{m}^\dag  \hat{m}\,,\, \hat{m}^T \hat{m}^*\big]\, =\, 0.
\eeq
whence (\ref{Alin}) vanishes with this choice of $\cU_0$.
We emphasize that \refer{mm} does not restrict the parameters of
the Lagrangian in any way, but simply tells us how  the matrix $\cU_0$ is determined 
from $Y^\nu$ in order to reach the true vacuum of the one-loop effective potential.
For notational simplicity we will drop the hats on the mass parameters in the remainder. 

Remarkably, the explicit form of the matrix $\cU_0$ is thus not needed,
it is enough to simply impose the condition \refer{mm}. For instance, in 
the so-called Casas-Ibarra parametrization  \cite{CA-IB} the redefined 
$\hat{Y}^\nu$ matrix has the form
\beq\label{Eq:Ynu-CI-form}
\hat{Y}^\nu=\frac{1}{v_H} U_\nu^* \sqrt{\bar{m}_\nu} R_{CI} \sqrt{\hat{M}},
\eeq
with a complex orthogonal $R_{CI}$ matrix and a unitary matrix $U_\nu$
(sometimes called PMNS matrix, being the neutrino analog of the CKM matrix);
furthermore, the diagonal matrix $\bar{m}_\nu$ of eigenmasses of light neutrinos
\beq
\bar{m}_\nu= {\rm diag}\big(\bar{m}_{\nu,1},\bar{m}_{\nu,2},\bar{m}_{\nu,3}\big).
\eeq
The general solution to \refer{mm} then requires (assuming ${\rm det}\, \bar{m}_\nu \neq0$)
\eq{\label{Eq:CI-real}
R_{CI}^*=R_{CI}.
}
Thus all phases of $\hat{Y}^\nu$ are contained in the PMNS  matrix.
{\em To simplify the notation we will from now on assume that the couplings have been appropriately redefined and drop the hats in all formulas.}

At quadratic order in $A$ there are eight contributions from the usual loop diagrams and two contributions from the tadpole diagrams with two external $A$ legs which endow the erstwhile Goldstone bosons with a (small) mass. At one loop the relevant contributions come from the diagrams depicted in Fig.~1 below.

\begin{center}
\includegraphics[scale=0.4]{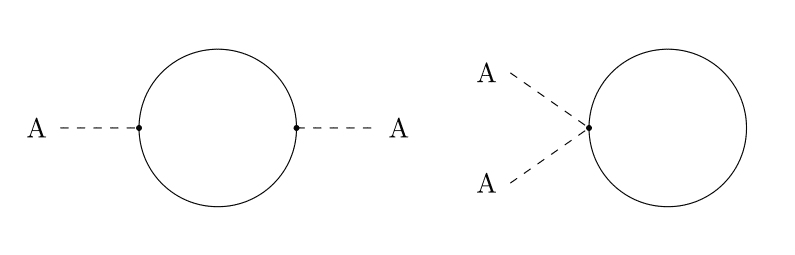} \\
{\scriptsize {\it Fig. 1.} Two types of diagrams that contribute to the quadratic terms in the potential for $A$. Every vertex couples to either $\nu^{i\al}N^j_\al$ or $\bar{\nu}^i_{\dot{\al}} \bar{N}^{j\dot{\al}}$; solid lines represent neutrino propagators from (\ref{expprops}).}
\end{center} 

Up to and including  $\cO(m^4)$ terms they are given by
\bea
(1)&=& \int \frac{d^4p}{(2\pi)^4} \Tr \bigg\{   \frac{-2}{p^2+M^2}m^\dag m A^2
          +\frac{2}{(p^2+M^2)^2}(m^\dag m)^2 A^2\nn\\
        && \qquad\qquad +\, \frac{-2M^2}{p^2(p^2+M^2)^2}m^\dag m A m^Tm^\star A\nn\\
        &&  \qquad\qquad +\, \frac{-M^2}{p^2(p^2+M^2)^2}\Big(\big[m^\dag m,A \big]\big[m^T m^\star,A\big]\Big)
        \bigg\},
\eea
and
\bea
(2)&=& \int \frac{d^4p}{(2\pi)^4} \Tr\bigg\{
     \frac{2}{p^2+M^2}m^\dag m A^2
          +\frac{-2}{(p^2+M^2)^2}(m^\dag m)^2 A^2\nn\\
 && \qquad\qquad +\, \frac{2M^2}{p^2(p^2+M^2)^2}m^\dag m A m^Tm^\star A\nn\\
  && \qquad\qquad  +\, \frac{-M^2}{p^2(p^2+M^2)^2}\Big(\big[m^\dag m,A\big]\big[m^T m^\star,A\big]\Big)
    \bigg\} \;\;.
\eea
Adding the two contributions we see that all the divergent terms cancel, so we are
left with a finite integral. Integrating over the momentum we arrive at the very simple and suggestive formula (now in terms of the dimensionful fields $G(x)$ introduced in (\ref{G}))
\beq\label{Amass}
\cL_{\rm eff} (A) \, \ni \, \frac{1}{8\pi^2 v_\phi^2} \Tr \Big( \big[ m^\dag  m, G \big]\big[ m^T m^* , G \big] \Big).
\eeq
Since the terms of order $\cO(m^2)$ cancel at one loop, and only terms $\cO(m^4)$ remain, one can worry that higher loop corrections can be more important than the contribution calculated above. However, there is a very simple argument showing
that the terms of order $\cO(m^2)$ will always cancel. Namely, if we focus on terms that do not contain derivatives of $A$, the only way $A$ can appear in the formula is through the exponential factor in Yukawa coupling (\ref{LYnewSimplified}). That means that the potential for $A$ can be calculated from the contributions to the vacuum energy by substituting $m \rightarrow m e^{-\ri A}$ in the formulae. Because the only structures of order $\cO(m^2)$ that can appear with the breaking pattern \refer{vevs} in the vacuum diagram are $m m^\dag$ and $m^* m^T$, which are invariant under this substitution, there will be no $\cO(m^2)$ terms in the potential for $A$, at any loop order. Terms containing $m m^T$ and $m^* m^\dag$ that potentially could provide contributions, will not appear because lepton number is conserved in the SM. The terms of order $\cO(m^4)$ can only appear because the allowed structure $m m^T m^* m^\dag$ is not invariant under this substitution. This also shows why commutator structures appear:
\begin{align}
m m^T m^* m^\dag &\rightarrow m e^{-2\ri A} m^T m^* e^{2\ri A} m^\dag = \\
&= m m^T m^* m^\dag -2\ri m [A,m^T m^*] m^\dag - 2m [A,[A,m^T m^*]] m^\dag +\dots \nn
\end{align}
If we had additional scalar fields like in (\ref{scpot1}), or any other mechanism for which 
$M_{ij}\sim\langle\phi_{ij}\rangle$ is not proportional to the identity matrix,
then the structures that can appear are more complicated. For example, instead of simple structure $m m^\dag$, we could have $m f(M) m^\dag$, with $f$ being some function of matrix $M$. Substituting $m \rightarrow m e^{-\ri A}$ now produces the following terms:
\begin{align}
m f(M) m^\dag &\rightarrow m e^{-\ri A} f(M) e^{\ri A} m^\dag = \\
&= m f(M) m^\dag -\ri m [A, f(M)] m^\dag +\dots \nn
\end{align}
Ultimately, those of the fields $A$ that do not commute with matrix $M$ will obtain mass terms of order $\cO(m^2)=\cO((\sqrt{m_\nu M})^2)$. For those that do commute however, this terms will vanish, and the leading contribution to their mass will come from (\ref{Amass}). 

The finiteness of the result (\ref{Amass}) is crucial,
and this is the sense in which the approximate SU(3)$_N$ symmetry `protects'  the pseudo-Goldstone bosons from acquiring large masses. If there were divergences the pseudo-Goldstone masses would have to be fixed by some renormalisation procedure, and we could no longer claim that they are `naturally' small. We also note that (\ref{Amass}) vanishes for diagonal $G(x)$,
hence two of the Goldstone bosons remain massless at this order (but not beyond).
As (\ref{Amass}) shows, the mass values are slightly smaller than the (light) neutrino masses.
Likewise, the part in $A$  proportional to the unit matrix  would drop out in this formula, and
the associated Goldstone boson would thus remain massless (but we note  that this
formula is anyway not  directly applicable to the Majoron $\Ca(x)$ as this field drops out from
the vertex (\ref{LYnew}) after re-defining all SM fermions).

\subsection{Pseudo-Goldstone couplings}

The pseudo-Goldstone particles couple, via the Yukawa interaction (\ref{L'}), to the usual (`non-sterile')  SM particles. Because these couplings receive non-vanishing contributions only at higher orders in the loop expansion they are naturally small, with calculable coefficients \cite{LMN}, and this fact makes them obvious Dark Matter candidates. In this subsection we briefly discuss some of the possible couplings, in particular the couplings to neutrinos and photons. These are not only relevant to the question which pseudo-Goldstone excitations can survive to the present epoch and hence serve as viable dark matter candidates, but also to
the question whether their decays can be observed in principle. %
The decays of these pseudo-Goldstone bosons into other lighter pseudo-Goldstone bosons are strongly suppressed.

The first point to note is that our pseudo-Goldstone bosons cannot decay into light
neutrinos because by (\ref{Amass}) their masses are generically below the light neutrino mass values.
This is crucial for them to be viable dark matter candidates, as otherwise they would have decayed long ago! However, they can decay into photons, with a calculable rate. This rate follows from an explicit calculation of the effective vertex
\beq
\mathfrak{L}_{\rm eff} \sim \frac1{v_\phi M^4}\
\sum_i  g_{A\ga\ga,i}\big( m^*\big[ m^\dag m , A \big]m^T \big)_{ii} F_{\mu\nu}F^{\mu\nu},
\label{AFFcoupling}
\eeq
where $g_{A\ga\ga,i}$ is of the order of $10^{-4}$ for $M\sim 200$ GeV. Observe that the coefficient $g_{A\ga\ga\, i}$ depends on the family index $i$ via the mass of
the associated lepton $m_i \equiv (m_e, m_\mu, m_\tau)$, otherwise this
term would vanish altogether with our minimization condition from (\ref{mm}).
As a consequence the result depends on the differences between the
contributions from different leptons. Even without taking this into account the
effective decay rate is extremely small
\beq
\Ga_{A\ga\ga}\,\sim\, \frac{g_{a\ga\ga}^2m_\nu^2}{8\pi M^2}\frac{m_A^3}{v_\phi^2}\,\ll  \,
10^{-42} \eV .
\eeq
This is many orders of magnitude less than the Hubble parameter ($H_0^{-1}\sim 10^{-32}$ eV). Therefore we conclude that these pseudo-Goldstone bosons are stable.

The result (\ref{AFFcoupling}) may also be important for axion searches (see \cite{OSQAR}).
However, for the present model with only a sterile scalar sextet, the effective coupling is of the order of
\beq
\frac{1}{f_\ga}\sim \frac {m_\nu^2}{M^3}\sim 10^{-24}\ {\rm GeV}^{-1},
\eeq
and thus far beyond the reach of current experiments. However, this situation
may well change in the presence of more complicated scalar sectors, such as 
the one discussed in the Appendix:
if the eigenvalues of the mass matrix $M$ of the heavy neutrinos were different from each other, the coupling would be of order of $\frac{m_\nu}{M^2}\sim 10^{-13}\ {\rm GeV}^{-1}$.
This  value would still pose a challenge, but could be much closer to experimental verification.

We would also like to emphasize that the present model in principle allows not only
for couplings of the type (\ref{AFFcoupling}), but also for axionic couplings
$\propto A F\tilde F$, such that there can appear effective couplings
\beq
\mL_{\rm eff} \,\ni \, \frac12 a A_0\, ({\bf E}^2 - {\bf B^2}) \, + \, b A_0 \, {\bf E}\cdot {\bf B},
\eeq
with computable small coefficients $a$ and $b$, and $A_0\equiv \sum_{i,j} c_{ij} A_{ij}$
a certain linear combination of the pseudo-Goldstone bosons.

In principle the pseudo-Goldstone bosons also couple to gluons,  again with computable coefficients. As before the coupling need not be purely axionic. Not unexpectedly, the coupling
turns out to be extremely small: for the present model it is
proportional to ( see \cite{Lat} for a derivation)
\beq\label{AGG}
\mL_{\rm eff} \,\ni \,  \frac{\al_W^2 y_M}{8\pi^2 M_W^4 v_\phi} \Tr \left( m\big[ M^\dag M , A \big]m^\dagger\right) \left[ \frac{\al_s}{4\pi}\Tr (G^{\mu\nu}\tilde{G}_{\mu\nu}) \right],
 \eeq
in lowest order (involving several three-loop diagrams as in \cite{LMN}).  This expression vanishes if the matrix $M^\dagger M$ is proportional to unity,  in which case one would have to go to the next order to obtain a non-vanishing result. However, it is possible to obtain a non-vanishing result already at this order with a more complicated scalar sector.

\section{Constraints and predictions}

Any of the following three observations would immediately falsify the model:
\begin{itemize}
\item Discovery of a genuinely new mass scale (proton decay, WIMPs, {\it etc.});
\item Detection of new fundamental spin-$\frac12$ degrees of freedom;
\item Detection of {\em non-sterile charged} scalar degrees of freedom, as predicted by 
 two-doublet models or all models of low energy supersymmetry (squarks, sleptons, etc.).
\end{itemize}

If none of the above shows up in the near future the model presented in this paper (or some modified version thereof) can be considered as an alternative.

The first test of the proposed scenario is of course whether it is possible at all to arrange the parameters such that all the conditions and constraints imposed by observations can be simultaneously satisfied in such a way that no large intermediate scales are needed, and the subset of couplings already known from the  SM agrees with the ones computed in our model.
We now list the conditions that will have to be met for our scenario to be consistent and compatible with what has been observed so far.

\subsection{Perturbative Consistency}
Scalar fields are usually accompanied by quadratic divergences, which
are generally viewed as posing a fine tuning challenge.  With several new scalar
fields beyond the SM scalar sector we have to address this issue. The desired
cancellation of quadratic divergences is one of the main motivations for `going
supersymmetric', but we will here follow a different, and more economical strategy,
by imposing the cancellation of quadratic divergences directly in terms of bare parameters at the Planck scale \cite{SBCS}. The underlying assumption here is that at this scale a proper and finite theory of quantum gravity (not necessarily a space-time based quantum field theory) `takes over' that will explain the cancellation in terms of some as yet unknown symmetry (different from low energy $N=1$ supersymmetry).
The corresponding conditions were already evaluated for a simpler model in \cite{SBCS}, where it was shown that a realistic window could be found for the couplings. This analysis can be generalized to the present case.

In addition we require that none of the couplings should exhibit Landau
poles over the whole range of energies from the electroweak scale to
the Planck scale. Likewise, there should be no instabilities (in the form
of lower unboundedness) of the effective potential over this range. Realizing
this assumption in the concrete model at hand shows that the putative
instability of the Higgs potential in the (un-extended) SM
(see e.g. \cite{Degrassi,Buttazzo,HKO,Branchina}) can be avoided altogether. Obviously, these requirements
lead to strong restrictions on the couplings, and it is one of the main challenges
whether these can be met with our other assumptions.

As explained in \cite{SBCS} for each scalar we impose the vanishing of the quadratic divergence associated with this scalar at the Planck mass, and then evolve back to
the electroweak scale, matching the couplings to the electroweak couplings as
far as they are known. For the investigations of the scale dependence of the couplings at one loop we need the coefficients in front of the quadratic divergences; they read
\bea
\label{Eq:QuadDivCoeff}
f_{H}&=&\frac{9 }{4}g_w^2+\frac{3 }{4}g_y^2+6 \lambda _1+12 \lambda_3-6 y_t^2,\nn\\[2mm]
f_\phi&=& 14 \lambda _2+4 \lambda _3+8 \lambda _4-|y_M|^2.
\eea
Non-zero values of  $Y^\nu$ do not produce additional
quadratic divergences at one loop, except for a negligible contribution to $f_H $.
At one loop the $\beta$-functions do not depend on the renormalization scheme, and can be
deduced from the general expressions given in \cite{JO};  they are ($\tilde\beta\equiv 16\pi^2 \beta$)
\bea\label{beta}
\tilde\beta_{g_w} &=& -\frac{19}{6}g_w^3\;, \qquad\quad
\tilde\beta_{g_y}=\frac{41}{6}g_y^3, \qquad\quad
\tilde\beta_{g_s} =-7 g_s^3, \nn\\[2mm]
\tilde\beta_{y_t} &=& y_t\left\{ \frac{9}{2}y_t^2
-8g_s^2  -\frac{9}{4}g_w^2
-\frac{17}{12}g_y^2 \right\},\qquad\quad
\tilde\beta_{y_M}\! = \frac{5}{2}y_M|y_M|^2,
\nn\\[2mm]
\tilde\beta_{\la_1}&=&\frac{3}{8} \left(3 g_w^4+2 g_w^2 g_y^2+g_y^4\right) -6y_t^4 -3\left(3 g_w^2+g_y^2-4 y_t^2\right) \lambda _1+\nn\\
 &{}&+12 \left(2
   \lambda _1^2+2 \lambda _3^2\right),\nn\\[2mm]
\tilde\beta_{\la_2}&=&40 \lambda _2^2+8 \lambda _3^2+6 \lambda _4^2+32 \lambda _2 \lambda _4+2\lambda_2 |y_M|^2,\nn\\[2mm]
\tilde\beta_{\la_3}&=&
\lambda _3\! \left[|y_M|^2+6 y_t^2-\frac{9 g_w^2}{2}-\frac{3
   g_y^2}{2}+12 \lambda _1+28 \lambda _2+8 \lambda _3+16
   \lambda _4\right],\nn\\[2mm]
\tilde\beta_{\la_4}&=&
22 \lambda _4^2+\left(2 |y_M|^2+24 \lambda _2\right) \lambda
   _4-|y_M|^4,
\eea
and
\bea
\tilde{\beta}_{m_1^2} &=&
m_1^2 \left(12 \lambda _1-\frac{3}{2} \left(3 g_w^2+g_y^2\right)
+6 y_t^2\right)+24 \lambda _3 m_2^2,\nn\\[2mm]
\tilde{\beta}_{m_2^2} &=& 8 \lambda _3 m_1^2
+m_2^2 \left(28 \lambda _2+16 \lambda _4+|y_M|^2\right).
\eea
Anomalous dimensions (in the Landau gauge) can be derived from the above expressions and the effective 
potential given below
\beq
\gamma_\phi=\frac{1}{32 \pi ^2}|y_M|^2,\qquad
\gamma_H=-\frac{3}{64 \pi ^2} \left(3 g_w^2+g_y^2-4 y_t^2\right).
\eeq
We also refer to \cite{HKO} for an investigation of the scale dependence of $f_H$ in the (un-extended) Standard Model.

\subsection{Vacuum stability}

One of the important open issues for the SM concerns the stability of the electroweak 
vacuum. There are strong indications that this vacuum develops an instability around
$\sim 10^{11}\,$GeV when radiative corrections are taken into account 
\cite{Degrassi,Buttazzo,HKO,Branchina}. More specifically, the {\em RG improved} one-loop
potential $\mathcal{V}^{RGI}_{\rm eff}(H) \sim \lambda (\mu=H) H^4$ becomes negative when
the running coupling $\lambda(\mu=H)$ dips below zero, as it does for large field values
$H \sim 10^{11}\,$GeV. Remarkably, however, the potential fails to be positive by
very little, so one might hope that a `small' modification of the theory might remedy
the instability. We will
now argue that this is indeed the case for the present model.~\footnote{This stability
 requirement was already present in previous versions of the CSM \cite{MN,SBCS}.
 See also \cite{CC} for an alternative proposal how to stabilize the electroweak vacuum.}

To confirm that the point \refer{vevs} is indeed the global minimum  of the full effective 
potential we recall that we impose the conditions of positivity of the quartic potential 
(listed in section \ref{Sec:ScalarSec}) for all values of the RG scale $\mu$ between the electroweak and the Planck scale. In order to investigate this issue more carefully 
we note that for $Y^\nu=0$ the effective potential has an exact U$(3)$ symmetry,
and thus  reaches all its values on a submanifold parametrized by 
\beq
{H} = \frac{1}{\sqrt{2}}\left(\begin{array}{c} 0 \\[1mm]
                                                        \varphi_4  \end{array}\right), \qquad
\phi=
\frac{1}{\sqrt{2}}\left(\begin{array}{ccc} \varphi_1&0&0 \\[1mm]  0&\varphi_2&0 \\[1mm]   0&0&\varphi_3  \end{array}\right),
\eeq
with nonnegative parameters $\varphi_i$. Its explicit form (in the Landau gauge and the $\overline{MS}$ scheme of dimensional regularization) reads 
\bea \label{Veff}
\cV_{\rm eff}(\varphi)\,=\, \cV({H}\big(\varphi), \phi(\varphi)\big)
\,+\, \hbar\cV^{(1)}(\varphi) \,+\, \mathcal{O}(\hbar^2), 
\eea
with the tree-level potential given in \refer{scpot} and (we follow the notation of \cite{FJJ})
\bea
64\pi^2\,\cV^{(1)}(\varphi)
&=&
\phantom{-2}\sum_{i=1}^{13}S_i^2\!\left\{\ln\frac{S_i}{\mu^2}-\frac{3}{2}\right\}
+3\,\mathbb{G}^2\!\left\{\ln\frac{\mathbb{G}}{\mu^2}-\frac{3}{2}\right\}
+\nn\\
&{}&-2\sum_{i=1}^{3} N_i^2\!\left\{\ln\frac{N_i}{\mu^2}-\frac{3}{2}\right\}
-12\, T^2\!\left\{\ln\frac{T}{\mu^2}-\frac{3}{2}\right\}+\nn\\
&{}&+3\, Z^2\!\left\{\ln\frac{Z}{\mu^2}-\frac{5}{6}\right\}
+6\, W^2\!\left\{\ln\frac{W}{\mu^2}-\frac{5}{6}\right\},
\eea
where
\beq 
W=\frac{1}{4}g_w^2 \varphi_4^2,\quad
Z=\frac{1}{4}(g_w^2+g_y^2) \varphi_4^2,\quad
T=\frac{1}{2}y_t^2 \varphi_4^2,\quad
N_i=\frac{1}{2}|y_M|^2 \varphi_i^2,
\eeq
\beq
\mathbb{G}=\lambda _1 \varphi _4^2+\lambda _3 \left(\varphi _1^2+\varphi _2^2+\varphi _3^2\right)+m_1^2,
\eeq
and 
\beq
S_i=\lambda _4 \varphi _i^2+\lambda _3 \varphi _4^2+\lambda _2 \left(\varphi _1^2+\varphi _2^2+\varphi _3^2\right)+m_2^2,\qquad{\rm for\  } i=1,2,3,
\eeq
\beq
S_{4-9}=
-\lambda _4 \left(\varphi_k^2\pm\varphi _l \varphi_n\right)
+\lambda _3 \varphi _4^2
+\left(\lambda _2+\lambda _4\right) 
\left(\varphi_1^2+\varphi_2^2+\varphi _3^2\right)+m_2^2,
\eeq
with $k,l,n=1,2,3$ and $k\neq l\neq n \neq k$. 
Finally $S_{10}$-$S_{13}$ are eigenvalues of the following $4\times4$ matrix
\beq
\mathcal{S}=\left(
\begin{array}{cccc}
 D_1 & 2 \lambda _2 \varphi _1 \varphi _2 & 2 \lambda _2 \varphi _1 \varphi _3 & 2 \lambda _3 \varphi _1 \varphi _4 \\
 2 \lambda _2 \varphi _1 \varphi _2 & D_2 & 2 \lambda _2 \varphi _2 \varphi _3 & 2 \lambda _3 \varphi _2 \varphi _4 \\
 2 \lambda _2 \varphi _1 \varphi _3 & 2 \lambda _2 \varphi _2 \varphi _3 & D_3 & 2 \lambda _3 \varphi _3 \varphi _4 \\
 2 \lambda _3 \varphi _1 \varphi _4 & 2 \lambda _3 \varphi _2 \varphi _4 & 2 \lambda _3 \varphi _3 \varphi _4 & E \\
\end{array}
\right),
\eeq
where
\bea
D_i&=& \left(2 \lambda _2+3 \lambda _4\right) \varphi _i^2
+\lambda _3 \varphi _4^2
+\lambda _2 \left(\varphi _1^2+\varphi _2^2+\varphi _3^2\right)+m_2^2,
\nn\\
E&=&
3 \lambda _1 \varphi _4^2+\lambda _3 \left(\varphi _1^2+\varphi _2^2+\varphi _3^2\right)+m_1^2.
\eea

Typically, the \emph{unimproved} one-loop potential \refer{Veff} with 
$\mu=M_t\approx 173\,\GeV$ can exhibit  an instability below the 
Planck scale. However, this effect is spurious, as its origin is entirely due to 
large logarithms. Although the method of RG improved effective potentials 
$\ \mathcal{V}_{\rm eff}^{RGI}$ is not as powerful in the multifield case as in models with 
only {\em one} scalar field~\footnote{In particular, for multifield models with classical 
  conformal symmetry, instead of constructing $\ \mathcal{V}_{\rm eff}^{RGI}$ one 
  usually exploits the RG             invariance to determine the `best' value of the RG scale 
  (i.e. the one for which the tree-level potential has a flat direction), following Gildener 
  and   Weinberg \cite{GW}.}, we can nevertheless formulate an RG improved version 
by taking the field dependent `radial norm'
\beq
\mu^2(H,\phi) = 2\left\{H^\dagger\!H+\Tr(\phi^*\phi)\right\}=\sum_{i=1}^4\varphi_i^2
\equiv|\!|\varphi|\! |^2,
\eeq 
as the scale parameter in field space. Then one checks numerically that
 (the RG improved version of) the potential \refer{Veff} remains positive for large values of $|\!|\varphi|\! |$ in the range 
$$
10{\rm TeV}\,\lesssim \, |\!|\varphi|\!|\,\lesssim\,  M_{Pl},$$
(in particular this is true for all points in the Table). This is a strong indication
that the electroweak vacuum \refer{vevs} remains the global minimum over 
this whole range of energies. The apparent discrepancy between the
unimproved and the improved effective potential is the same as for the SM, where the 
unimproved one-loop effective potential likewise reaches the instability already for much 
smaller field values than the RG improved one.

\subsection{Dark Matter constraints}

We have already pointed out that the pseudo-Goldstone bosons of our model
are natural Dark Matter candidates. However, in order to verify that they are really viable
we need to check (1) whether they can be non-relativistic,
and (2) whether they can survive till the present epoch \cite{GoRu}.
As for the second requirement, we have already checked  that the pseudo-Goldstone cannot  decay into light neutrinos.
The decay rate into photons was found to be very small, and many orders of
magnitude smaller that the present Hubble parameter. Hence the pseudo-Goldstone
particles are indeed `stable'.

The first requirement can be satisfied if at the time of the electroweak phase transition, {\it i.e.} for temperatures around 100 GeV, the causally connected region is smaller than the inverse mass of the Dark Matter candidate.
This requirement comes from the fact that the potential for the scalar fields
started to be nonvanishing at the time of the electroweak transition. At that point, the phase fields start to oscillate coherently, and the fluctuations of smaller wavelength than the causal region are suppressed. To get a rough estimate, we note that the causally connected region at that time of the phase transition ($\sim 10^{-10}$ s) was about 0.01 m; expressed in mass units this corresponds to a mass bigger than about $10^{-4}$ eV. As we can see from the formula (\ref{Amass}) the masses of the pseudo-Goldstone bosons are not too much below the mass of the light neutrinos, so this requirement can be satisfied and they are naturally in a (small) window between $10^{-4}$ eV and the light neutrinos masses.

An equally important point concerns the abundance with which the Dark Matter
particles are produced, so as to arrive at the desired value $\Omega_{DM} \sim 0.3$. In order
to derive a very rough estimate we note that this requires (amongst other things)
not only a knowledge of the pseudo-Goldstone masses, but also of the effective
potential $V_{\rm eff}(G)$. All we know is that the latter must be a single-valued function
on the Goldstone manifold SU(3)$_N$/SO(3), cf. (\ref{MGoldstone}).
It is also clear from our foregoing considerations this potential is in principle calculable
via the determination of the effective higher point vertices of the pseudo-Goldstone fields. 
At one loop the effective potential in $G$ derives from
\beq
V_{\rm eff}(G) \,\propto \, \Tr \big( m e^{-iG/v_\phi} m^T m^* e^{iG/v_\phi} m^\dag \big),
\eeq
which yields the estimate
\beq\label{Vmax}
V_{\rm eff}^{max} \,\propto\,  \Tr (mm^Tm^* m^\dag)   \, \sim \, m_\nu^2 M^2,
\eeq
for the height of the potential. The contribution to $\Omega$ then follows from scaling
down the energy density of the pseudo-Goldstone particles
to the present epoch by means of the factor $(R_*/R_0)^3 \sim
(T_0/T_*)^3$ where $R_0$ ($T_0$) is the present radius (temperature) of the universe,
and $R_*$ ($T_*$) the radius (temperature) of the universe when the abundance is
produced. To estimate the latter, we observe that for $T> V_{\rm eff}^{max}$ we have
thermal equilibrium, and only for $T < V_{\rm eff}^{max} $ can the pseudo-Goldstone
particles start to be produced non-thermally by coherent oscillations. Therefore
setting $T_* = (V_{\rm eff}^{max})^{1/4}$ seems a reasonable choice; this gives
\beq
\Omega \;\sim \; \rho_{crit}^{-1} V_{\rm eff}^{max} \left(\frac{T_0}{T_*}\right)^3
   \; \sim \; \rho_{crit}^{-1} \left( V_{\rm eff}^{max}\right)^{1/4} T_0^3 \;\sim\;
   \rho_{crit}^{-1} \sqrt{m_\nu M} \, T_0^3.
\eeq
This is indeed an estimate that also gives about the right order of magnitude for standard
axions, with $V_{\rm eff}^{max} = \Lambda_{QCD}^4$. In our case, the result comes
out too small by two or three orders of magnitude. However, the above estimate is fraught with
several uncertainties, apart from the precise details of the production mechanism,
which may give rise to all kinds of `fudge factors'.  In particular, since there is
a `collective' of scalar fields involved in this process it is not clear whether there cannot exist new enhancement effects, similar to the resonant effects giving rise to leptogenesis
as in \cite{PU,PU2}.  Furthermore, a modification of the scalar sector along the lines
of section~2.2 might change the value of $V_{\rm eff}^{max}$, for instance 
replacing $m_\nu^2 M^2$ by $m_\nu M^3$ in (\ref{Vmax}) which would give the
desired number. So this issue clearly requires further and more detailed study.

\subsection{Leptogenesis}

An important feature of the present model is that it can account for the observed
matter-antimatter asymmetry ($\sim 10^{-10}$) in a fairly natural manner. 
Since the masses of right-chiral neutrinos are smaller than the usually quoted bound ($\gg 10^5$ TeV) we have to assume that the source of the asymmetry is {\em resonant leptogenesis} \cite{PU, PU2}. One of the necessary conditions for this mechanism to work is the approximate degeneracy of the masses of right-chiral neutrinos -- exactly as  obtained in our model.
The shift $\delta M$ induced by the Dirac-Yukawa term is naturally very small,
and turns out to be exactly of the magnitude required by the condition given in \cite{PU}:
\eq{
\de M\,\sim \, \Ga.
}
This is because, on the one hand, the decay rate of a massive neutrino in our model
is  $\Ga  \, \sim \, Y_\nu^2 M $. On the other hand, the mass splitting 
induced by the Dirac-Yukawa coupling is $\de M \,\sim \,  Y_\nu^2 M$;
the latter is caused by two sources -- the diagonalization of the neutrino mass matrix in the presence of the Dirac Yukawa term (\ref{LYnew}) and the RGE running of the Majorana-Yukawa couplings from $M_{PL}$ down to ${\rm TeV}$ scale of heavy neutrinos. It is important to emphasize that the condition $\de M\sim\Ga$ is thus very natural in our model, 
whereas it usually requires a certain amount of fine tuning, especially in GUT type models.

If we use the formula to estimate the baryonic asymmetry given in \cite{PU2} 
we get the correct asymmetry taking into account light neutrino data and assuming nonzero (but small) phases of the PMNS  matrix in Eq. \refer{Eq:Ynu-CI-form}. In our case, as we have already said in \refer{Eq:CI-real}, the Casas-Ibarra matrix has to be real, so that the PMNS phases are responsible for the leptogenesis.
For example, the points shown in Tab. \ref{Tab:Points}, give $\eta_B\approx 6\times 10^{-10}$ with PMNS phases of order $10^{-3}$. We leave the details of this and other leptogenesis related calculations to a future publication.

\subsection{New scalar particles}

Because much of the new structure of the model is sterile, not many dramatic new effects are expected to be observable beyond the SM. Nevertheless, there are
distinctive signatures that are very specific to the present scenario, and
that can be easily used to discriminate it from other BSM scenarios. These
are mainly due to the mixing of the new scalars with (the $H_0$ component of)
the Higgs doublet induced by the potential (\ref{scpot}) with (\ref{vevs}).
From (\ref{hh'}) we immediately get he decomposition of $H_0$ in terms of
the mass eigenstates $h_0$ and $h'$
\bea
H_0&=& \cos\beta \, h_0 \, - \, \sin\beta \, h',
\label{hmixing}
\eea
whence the scattering amplitude would be well approximated by
\bea
\cA\;&\propto& \;
\frac{\cos^2 \beta}{p^2 + m_{h_0}^2 + i\cos^2\beta \,m_{h_0}\,\Gamma_{SM}(m^2_{h_0})} \; +
\nn\\[2mm]
&& \qquad\qquad
\, + \,
\frac{\sin^2\beta}{p^2 + m_{h'}^2 + i\sin^2\beta  \, m_{h'}\,\Gamma_{SM}(m^2_{h'})}.
\eea
The existing experimental data suggest that $|\cos\beta|$ should be close to 1, if
$h_0$ is to mimic the SM Higgs boson. The particle corresponding to $h'$ has not been observed yet. The mixing will thus induce interactions of this new mass eigenstates with SM particles. In particular the decay channels of the standard Higgs boson are also open to the new scalar excitations, possibly leading to a kind of `shadow Higgs' phenomenon, with decay amplitudes of approximately the same height but sharply reduced width \cite{MN325}. In addition, depending on the mass values of the new scalars there may be extra decay channels involving new scalars, and possibly even heavy neutrinos, leading to a broadening of the resonance curve.

The existence of new scalar degrees of freedom mixing with the standard Higgs boson
is the main {\em generic prediction} of the present model. It is a distinct signature that, though perhaps not so easy to confirm, can serve to discriminate the present model from other
scenarios, in particular supersymmetric and two-doublet models which inevitably
contain {\em non-sterile} scalars, or the $\nu$MSM model of \cite{Shap}, which does have a
sterile scalar (and also keV range `heavy' neutrinos), but absolutely nothing above $m_{h_0}$
in the TeV range. Thanks to the mixing the new scalar(s) may eventually be seen at LHC, but the actual discovery potential for discovery depends, of course, on their masses, mixing angles etc. The mixing would also lead to a slight diminution in the decay width of the SM Higgs boson that can be measured in future precision tests at the Higgs resonance.

\subsection{Numerical analysis}\label{SubSec:Na}

We conclude this section by giving some numerical data which show
that there exists a wide range of points in parameter space with the following properties:
\begin{itemize}
\item The quartic potential $\mathcal{V}^{quart}({H}, \phi,\mu)$ is positive definite for all renormalization scales $\mu$ between $M_t$ and the $M_{Pl}$, while all dimensionless
couplings $c(\mu)=(\lambda(\mu),g(\mu),y(\mu))$ remain perturbative in this range (i.e. $|c(\mu)|<4$ in our normalization conventions);
\item the coefficients functions $f_i$ of the quadratic divergences
defined in (\ref{Eq:QuadDivCoeff})  vanish at the Planck scale;
\item For $\mu=M_t$ there exists a stationary point of the type \refer{vevs}, with $v_H\approx 174\,{\rm GeV}$, which is the {\em global} minimum of the potential \refer{scpot}; moreover, the
SM-like Higgs particle can be arranged to have $M_{h_0}=125\, {\rm GeV}$ such that
$|t_\beta|<0.3$, cf. \refer{hmixing};
\item There exists a matrix $Y^\nu$ consistent with both Dashen's conditions and light neutrino data that yields $\eta_B\approx 6\times 10^{-10}$ as well as a positive semi-definite pseudo-Goldstone boson mass matrix corresponding to \refer{Amass}.
\end{itemize}

Some representative numerical  examples are listed in Tab. \ref{Tab:Points} with $y_M=y_M(\mu=M_t)$. We also show  there decay width of the `shadow Higgs' $h'$ and the branching
ratios of $h_0$ and $h'$ into `old particles' ($\equiv OP$),  i.e. particles  discovered prior to 2012. All points have $M_{h_0}=125 \GeV$ and $v_H=174\GeV$.

\begin{table}[h]
\caption{Example points (all dimensional parameters are given in $\GeV$)}   \label{Tab:Points}
\begin{tabular}{|c|c|c|c|c|c|c|c|}
\hline
 $|y_M|$& $M_N $& $M_{h'}{}$ & $M_{R}{}$ &
   $t_{\beta }$ & $\Gamma_{h'}$ & ${BR(h' \to OP)}$ &
   ${BR(h_0\to OP)}$ \\\hline\hline
0.56 & 545 & 378 & 424 & -0.3 & 3.1 & 0.59 & 0.69 \\\hline
 0.54 & 520 & 378 & 360 & -0.3 & 3.1 & 0.59 & 0.68 \\\hline
 0.75 & 1341 & 511 & 1550 & 0.25 & 6.2 & 0.73 & 0.91 \\\hline
 0.75 & 2732 & 658 & 3170 & -0.16 & 5.9 & 0.74 & 0.99 \\\hline
 0.82 & 2500 & 834 & 2925 & 0.15 & 10.9 & 0.74 & 0.98\\\hline
\end{tabular}
\end{table}

\section{Gauging $(B\! - \!L)$}
While the consistency of the model introduced in the previous sections does not require 
any further modifications some of them seem self-evident. For instance, by further
enlarging the scalar sector, there can appear additional `shadow Higgs bosons'.
In the appendix we present one such example with a new scalar triplet $\xi$
in the fundamental representation of SU(3)$_N$, which is also 
in complete agreement with our basic assumptions. 

A more important (and perhaps also more plausible) possibility follows from  the 
cancellation of $(B\!-\!L)$ anomalies, a fact that is widely viewed as an indication that
this symmetry should be gauged  (see in particular \cite{CW5,CWa} for recent work 
in this direction in the context of conformal invariance). Thus one can enquire under what conditions gauging U(1)$_{B-L}$ would be consistent both with our assumptions and existing experimental bounds. 
The associated U(1)$_{B-L}$ gauge boson (\emph {alias} $B'$ boson) would then 
also appear in the scalar kinetic terms
\beq
\mL_{\rm kin} =-\Tr\!\!\left[ (\pa^\mu \phi^*+2ig_x B'^\mu\phi^*)(\pa_\mu \phi-2ig_x B'_\mu\phi)\right]
\, +\,  \cdots .
\eeq
From these and the expectation values (\ref{vevs}) we immediately deduce the
mass of the $B'$-boson
\beq
m^2_{B'} \,=\, 24 g_x^2 v_\phi^2 \, .
\eeq
This simple picture is complicated by the kinetic mixing of $B'$ with U(1)$_Y$ gauge boson $B$. The mixing can be equivalently described as a modification of covariant derivatives (with standard, diagonal kinetic terms of gauge boson) \cite{Chank}. For an arbitrary matter 
field we have (with $X\equiv B\!-\!L$)
\eq{\nn
D_\mu=\partial_\mu+{\rm i}\left[ 
 g_s\mathcal{T}_i\, G^i_\mu+g_w T_a\, W^a_\mu
+g_y Y B_\mu 
+(g_x X + g_m Y) B'_\mu\right],
}    
with generators  $\mathcal{T}_i,\  {T}_a$ for SU(3)$_C\times$SU(2)$_{W}$. The above form is invariant under RGE with non-standard `rotating' anomalous dimensions of gauge fields \cite{Chank}. However, the condition $g_m=0$ is {\em not} -- even if we start at the electroweak scale with pure $(B\!-\!L)$ gauge theory, an admixture of $Y$ is always generated in the RG flow. A non-zero value of $g_m$ produces non-diagonal elements of the (tree-level) mass matrix; 
in terms of mass-eigenstates we have
\eq{\nn
B'_\mu=\sin\zeta\, Z_\mu + \cos\zeta\, Z'_\mu,
}
where $Z_\mu$ is the SM-like $Z$ boson. In the analogous decomposition of $B_\mu$ and $W_\mu^3$ the photon field appears in addition to $Z_\mu$ and $Z'_\mu$.    
Preliminary checks show that one can find a range in parameter space consistent with the counterparts of conditions summarized in Sec. \ref{SubSec:Na}, as well as the LEP 
limits \cite{LEP1,LEP2} 
\eq{
|\zeta|\lesssim 10^{-3},\qquad \frac{M_{Z'}}{g_x}>7\, {\rm TeV}.
}
In our case typical values of the $Z'$ mass are below 10 {\rm TeV}. While the appearance of a $Z'$ gauge boson in this range would seem
difficult to reconcile with a GUT-type scenario, it would constitute clear
evidence for the present scheme! We also emphasize that the `pure $(B\!-\!L)$' model, defined 
by $g_m=0$ at the electroweak scale, is consistent with our conditions, and in particular
with the modified implementation of conformal symmetry (i.e. vanishing of the coefficients in front of quadratic divergences at the Planck scale). By contrast, the minimal 
`pure $(B\!-\!L)$' model is incompatible with vacuum stability if the symmetry is broken  
by means of the Coleman-Weinberg mechanism \cite{CWa}.   

\section{SU(3)$_N$ symmetry vs. quantum gravity?}
Finally, we would like to comment on one issue that concerns the eventual embedding 
of the present model into a UV complete theory of quantum gravity.\footnote{We
  would like to thank the anonymous referee for raising this point.}  Quantum gravity is widely
believed to be incompatible with {\em global} symmetries, whence only local (gauge)
symmetries are expected to survive in a `final' theory, and one might therefore worry
about possible implications of this folklore theorem for the present model. The argument
against global symmetries is basically related to the evaporation of black holes.
If there were conserved charges associated to global symmetries, these charges, when 
dropped into a black hole, would either `disappear' in  violation of charge conservation,
or otherwise, if the charges are really conserved, prevent black holes from decaying 
completely, necessarily leaving charged remnants. Since the initial black hole can 
in principle have an infinite number of charge quantum numbers, and since the 
associated objects would all look indistinguishably like Schwarzschild black holes
(this is where the absence or presence of gauge interactions makes all the 
difference), one would thus run into a potential conflict with black hole
entropy bounds \cite{CEB1,CEB2}. However, apart from the fact that black hole evaporation, 
and in particular its suspected {\em unitary} description, is still far from understood, we 
can proffer the following `physics proof'  that the present model evades such putative trouble. 
First of all, the SU(3)$_N$ {\em is} broken, both spontaneously and explicitly. Secondly,
this symmetry is very much in the spirit of the SU(3)$_L\times$SU(3)$_R$ flavor symmetry of the old quark model:
there as well, one has explicit as well as spontaneous symmetry breaking, with 
the pions as the pseudo-Goldstone bosons. Just like our SU(3)$_N$, the flavor symmetry  
looks like an {\em exact} global symmetry when viewed from the Planck scale, but there
is absolutely no evidence from meson physics that quantum gravity effects or black hole
evaporation modify or invalidate  the `naive' predictions of the model.
 
In fact, this argument can be made slightly more quantitative if one invokes wormholes
as the source of symmetry breaking  (as wormholes may `swallow'  global charges). While the relevant calculations are highly model dependent, one can safely assume 
that symmetry breaking effects are generically suppressed in the gravitational path
integral by a factor $e^{-\mathcal{S}} \sim f_0\slash M_P$,
where $f_0$ is the scale of symmetry breaking \cite{Wormholes}. From this estimate, 
effects of wormholes are indeed potentially relevant for axion phenomenology 
and the role of the Peccei-Quinn U(1) symmetry \cite{Kam} because the scale is 
$f_0 \ge 10^{12}\text{ GeV}$, as conventionally assumed.  By contrast, for our SU(3)$_N$, the symmetry breaking scale $f_0$ is of order $10^3 \text{ GeV}$. Assuming that the suppression factor equals $f_0\slash M_P$ we can neglect effects of gravity in comparison with those caused by $Y_\nu$, whence the potential corrections from quantum gravity to our predictions
are completely negligible.

\section{Conclusions}
We have proposed an extension of the Standard Model based on a new approximate 
SU(3)$_N$ symmetry acting only on right-chiral neutrinos and the new sterile scalars,
under which all SM fields are neutral. 
We have shown that SU(3)$_N$ symmetry breaking pattern naturally leads to a degeneracy of heavy neutrino masses and thus to resonant leptogenesis. Moreover, 
the masses and couplings of the resulting pseudo-Goldstone bosons make them viable Dark Matter candidates. At the same time the model is perturbative up to the Planck scale and the electroweak vacuum remains stable. The possibility of gauging $B-L$ symmetry as well as further extension of the scalar sector were also discussed.\\ 
\indent The main message of this paper is therefore that there may exist a (potentially  rich) sector of  `sterile' scalar particles not far above the electroweak scale  that would manifest itself chiefly through the mixing with the SM Higgs boson and the appearance of {\em narrow} resonances in the TeV range or below. This would be the main observable consequence of the present work.

\vspace{0.2cm}
\noindent{\bf {Acknowledgments:}}
We would like to thank Valery Rubakov for discussions.
A.Lewandowski and K.A.M. thank the AEI for hospitality
and support. A.Lewandowski and K.A.M. were supported by the Polish NCN grant DEC-2013/11/B/ST2/04046.

\appendix

\section{More sterile scalars?}

Given the fact that many approaches to unification and quantum gravity come with
an abundance of scalar fields it is entirely conceivable that there exists an even
larger sector of scalar fields, and in this sense our model is just the simplest
example. As one further example,  we briefly discuss in this appendix an extension of the model obtained by introducing a complex scalar triplet $\xi_i$ transforming as a $\bf 3$ under SU(3)$_N$, and how the presence of such an extra field would modify
the vacuum structure and  other aspects of the model. One new feature here is that
$\xi_i$ is even `more sterile'  than $\phi_{ij}$ in that not only it does not directly couple to SM particles (like $\phi_{ij}$), but cannot even couple to right-chiral neutrinos if we insist on renormalizability. As a consequence the associated new  pseudo-Goldstone excitations are even more weakly coupled to SM matter than those coming from $\phi_{ij}$.

With the extra triplet $\xi_i$, the most general renormalizable and U(3) symmetric scalar field potential reads
\bea
\cV({H}, \phi,\xi) &=&  m_1^2 \, {H}^\dag{H} + m_2^2 \, \Tr\!(\phi  \phi^*)
+ m_3^2 \xi^\dagger\xi +(m_4 \xi^\dagger\phi\xi^*+{\rm h.c})    \nn\\[2mm]
 && \!\!\!\!\!\!\!\!\!\!\!\!\!\!\!\!\!\!\!\!\!\!\!\!
 + \la_1 \,({H}^\dag{H})^2  +\,  2 \la_3 \, ({H}^\dag{H}) \Tr\!( \phi \phi^*)
 +  \la_2  \left[\Tr\!(\phi\phi^*)\right]^2 + \la_4 \, \Tr\! (\phi \phi^* \phi \phi^*)
 \nn\\[2mm]
 && \!\!\!\!\!\!\!\!\!\!\!\!\!\!\!\!\!\!\!\!\!\!\!\!
+ \la_5\,\xi^\dagger\phi\phi^*\xi
+2\la_6 \, {H}^\dag {H} \, \xi^\dagger\xi
+2 \la_7 \, \xi^\dagger\xi \,\Tr\!( \phi \phi^*  )
+\la_8 \, (\xi^\dagger\xi)^2,
\label{scpot1}
\eea
where all coefficient are real except for $m_4$ (traces are over family indices). This potential is manifestly invariant under
\beq\label{U3Nxi}
\phi (x) \quad \ra \quad U \phi(x) U^T \; , \quad
\xi (x) \quad \ra \quad U \xi(x),
\quad\quad\quad U\in {\rm U}(3).\\
\eeq
One point to note is that with $\xi_i$ one can easily arrange for `anisotropic'
expectation values $\langle\phi_{ij}\rangle$  not proportional to the unit matrix.
As before there exists a range of parameters for which the
global minimum of the potential takes the form
\beq\label{vevsxi}
\langle \xi\rangle =    \mathcal{U}_0\!
\left(\begin{array}{c} 0 \\[1mm]  0 \\[1mm]
                                                        e^{i \alpha} v_\xi \end{array}\right),\quad
\langle {H} \rangle = \left(\begin{array}{c} 0 \\[1mm]
                                                        v_H  \end{array}\right), \quad
\langle \phi \rangle = \mathcal{U}_0\! \left(\begin{array}{ccc} v_1&0&0 \\[1mm]  0&v_1&0 \\[1mm]   0&0&v_2  \end{array}\right)\!\mathcal{U}_0^T,
\eeq
with  positive parameters $v_\xi, v_H, v_1$, $v_2 \, ( \neq v_1)$, the phase $\alpha$ fixed by $\arg(m_4)$ and the vacuum alignment matrix $\mathcal{U}_0$ is of the same origin as before.
The important new feature due to the presence of $\xi_i$  is the special form of the matrix $\langle \phi_{ij} \rangle$, with the equality of the first
two diagonal entries being due to the fact that the expectation value $\langle\xi_i\rangle$ singles out one particular  direction in family space, thus also lifting the degeneracy in the heavy neutrino mass matrix obtained from (\ref{vevs}).

Because the residual symmetry of (\ref{vevsxi}) is SO(2), and the manifold of Goldstone bosons is the coset
\beq\label{MGoldstonexi}
{\cal M} = {\rm U}(3)\big/ {\rm SO}(2),
\eeq
whence there are now altogether eight (pseudo-)Goldstone bosons. These can be
parametrized as
\bea\label{ReDef1xi}
\phi (x)&=&  \cU_0\, e^{\ri A(x)} \, \tilde{\phi}(x) \, e^{\ri A(x)^T} \, \cU_0^T \nn\\
\xi(x) &=&   \cU_0\,  e^{\ri A(x)} \, \tilde{\xi}(x),
\eea
with
\beq\label{Eq:A-decompxi}
A(x)\equiv \sum_{a}\!{}' A_a(x) \la^a,
\eeq
and where the sum runs over those generators
$\la^a$ (now including $\lambda^9\equiv\mathbbm{1}$) that are spontaneously broken by vacuum \refer{vevsxi}.

The analysis of the vacuum structure is now more cumbersome than before.
Expanding $\tilde{\phi}(x)$ and $\tilde{\xi}(x)$ about the vacuum
expectation values (\ref{vevsxi})
\beq
\tilde{\phi}_{ij}(x)=\langle\phi_{ij} \rangle+ {\phi}'_{ij} (x),
\qquad
\tilde{\xi}_i(x)=\langle\xi_i\rangle+  {\xi}'_i (x),
\eeq
we have to ensure that the quantum fluctuations $\phi'_{ij}(x)$ and $\xi'_i(x)$ do not contain Goldstone bosons, as the latter are to be absorbed into $\cU(x)$. In other words, the fields
$\phi'_{ij}$ and $\xi'_i$ should only contain the ten heavy non-Goldstone modes.
This is ensured by imposing the condition (see \cite{Weinberg}, chapter 19)
\beq\label{Eq:WeinCond}
{\rm Im }\Big\{ \xi'(x)^\dagger \lambda^a \langle\xi\rangle
+ \Tr\!\big[ \phi'(x)^\dagger \left\{(\lambda^a \otimes \mathbbm{1} + \mathbbm{1}\otimes \la^a)
\langle\phi\rangle\right\}\big]\Big\}=0,\ \ \forall  {a\in\{1,\ldots,9\}}.
\eeq
As before the main observable effects are due to the mixing between the SM-like 
Higgs boson and the additional scalars, but there now appear three narrow resonances above the already discovered Higgs boson. Once again there exists a wide range of parameters for which 
the analogs of the conditions listed in Sec. \ref{SubSec:Na} are obeyed.

\end{document}